\documentstyle[preprint,floats,prd,aps]{revtex}
\input epsf.tex
\def\DESepsf(#1 width #2){\epsfxsize=#2 \epsfbox{#1}}
\begin{document}
\preprint{\vbox {\hbox{OCHA-PP-89}}}  
\draft
\title{Combined $B\to X_s\psi$ and $B\to X_s\eta_c$ decays as a  test of 
factorization} \author{Mohammad R. Ahmady
\footnote{JSPS Fellow} and Emi Kou}\address{Department of Physics \\
Ochanomizu University \\
1-1 Otsuka 2, Bunkyo-ku,Tokyo 112, Japan}

\date{January 1997}
\maketitle
\begin{abstract}
We calculate the inclusive decays $B\to X_s\psi$ and $B\to X_s\eta_c$ 
using factorization assumption. To investigate the bound state effect 
of the decaying B meson in these inclusive decays we take into account 
the motion of the $b$ quark using a Gaussian momentum distribution 
model.  The resulting correction to free quark decay approximation is 
around $\% 6$ at most.  Utilizing a potential model evaluation of the ratio 
of the decay constants $f^2_{\eta_c}/f^2_\psi$, it is 
shown that the ratio $R=\Gamma (B\to X_s\eta_c )/\Gamma (B\to X_s\psi )$ 
can be used as a possible test 
of factorization assumption.  

\end{abstract}
%
\newpage
Exclusive and inclusive nonleptonic B decays to charmonium states are of 
special interest 
theoretically and experimentally.  These decay channels, among other things, 
provide a powerful testing ground for color suppression and factorization 
in hadronic B decays\cite{fact}.  At the same time, exclusive modes of 
two body 
$B$ decay into $K$ meson resonances and charmonium states can provide an 
alternative examination of models for the treatment of hadronic form 
factors\cite{hform}.

In this work, we focus on the inclusive two body decays $B\to X_s\psi$ 
and $B\to X_s\eta_c$, where $X_s$ is a final state hadron containing a 
strange quark.  There is no experimental data on the latter decay at this 
time.  However, as we point out, the eventual measurement of this 
inclusive decay channel can be used to test the validity of the 
factorization assumption in nonleptonic B decays.  In fact, one can show 
that the ratio $R=\Gamma (B\to X_s\eta_c )/\Gamma (B\to X_s\psi )$, 
calculated by using factorization, is independent of the QCD corrections 
and the scale ambiguity of the Wilson coefficients.  In this context, $R$ 
depends only on the ratio of the decay constants $f_{\eta_c}/f_\psi$ 
for which we use an improved estimate obtained in a previous work\cite{prd95}.

The inclusive decays $B\to X_s\psi (\eta_c)$ are usually approximated 
with the free quark decays $b\to s\psi (\eta_c)$.  To improve upon, in the 
present paper, we  
estimate the correction to this approximation by taking into account the 
motion of the b quark inside the B meson.  For this purpose, we use a one 
parameter Gaussian momentum distribution for the $b$ quark which has 
previously been applied to inclusive 
semileptonic\cite{accmm}, rare dileptonic\cite{ahhm} and 
nonleptonic B decays \cite{pps}(commonly known as ACCMM model in the 
literature).  We present results for a range of 
the model parameter obtained from fits to experimental data.  

Neglacting penguin operators, the relevant effective Hamiltonian for $B\to 
X_s\psi (\eta_c)$ can be written as:
\begin{equation}
\displaystyle
H_{eff}=\displaystyle\frac{G_F}{\sqrt{2}}V_{cb}^*V_{cs}
\begin{array}[t]{l}\displaystyle \left [C_1(\mu ){\bar c}^i\gamma^\mu 
(1-\gamma_5)b^i{\bar s}^j\gamma_\mu (1-\gamma_5)c^j \right . \\
\displaystyle \left .+C_2(\mu ){\bar 
s}^i\gamma^\mu 
(1-\gamma_5)b^i{\bar c}^j\gamma_\mu (1-\gamma_5)c^j \right ] +H.C. \; ,
\end{array}
\end{equation}
where $i$ and $j$ are color indices and $C_1(\mu )$ and $C_2(\mu )$ are QCD 
improved 
Wilson coefficients.  One then can use a Fierz transformation to write 
(1) in the following form:
\begin{equation}
H_{eff}=\displaystyle \frac{G_F}{\sqrt{2}}V_{cb}^*V_{cs}
\begin{array}[t]{l} \left [ \left  ( C_2(\mu ) + \frac{1}{3}C_1(\mu ) 
\right ) 
                 {\bar s}^i\gamma^\mu (1-\gamma_5)b^i
                 {\bar c}^j\gamma_\mu (1-\gamma_5)c^j \right .\\  
                 \displaystyle\;\;  + \left . C_1(\mu ){\bar s}^i\gamma^\mu 
                 (1-\gamma_5)T^a_{in}b^n{\bar c}^j\gamma_
                 \mu (1-\gamma_5)T^a_{jm}c^m \right ] +H.C. \; ,
\end{array}
\end{equation}
where $T^a\; (a=1..8)$ are generators of $SU(3)_{\rm color}$.  We note 
that the first term in (2) is the product of two color singlet currents 
but the second term consists of two color octet currents.  In the 
factorization assumption, only the color singlet quark current 
contributes to the production of colorless $c\bar c$ final state 
$\psi$ and $\eta_c$.  Using the definition for the decay constant 
($f_\psi$) for the vector meson $\psi$:
\begin{equation}
f_\psi \epsilon^\mu =<0|\bar c \gamma^\mu c|\psi > \; ,
\end{equation}
($\epsilon^\mu$ is the polarization vector of $\psi$) the effective 
Hamiltonian for $B\to X_s\psi$ is obtained as follows:
\begin{equation}
H_{eff}^{B\to X_s\psi}= 
Cf_\psi\bar s\gamma^\mu 
(1-\gamma_5)b\epsilon_\mu \; ,
\end{equation}
where
\begin{equation}
C= \frac{G_F}{\sqrt{2}}V_{cb}^*V_{cs}\left (C_2(\mu 
)+\frac{1}{3}C_1(\mu )\right )\; . \end{equation}
Similarly, utilizing the definition for the decay constant ($f_{\eta_c}$) 
for the psuedoscalar meson $\eta_c$:
\begin{equation}
f_{\eta_c} q^\mu =<0|\bar c \gamma^\mu\gamma_5 c|\eta_c (q) > \; ,
\end{equation}
results in the following effective Hamiltonian 
for $B\to X_s\eta_c$ decay: 
\begin{equation}
H_{eff}^{B\to X_s\eta_c}= 
Cf_{\eta_c}\bar s\gamma^\mu (1-\gamma_5)bq_\mu \; .
\end{equation}
The Wilson coefficients $C_1$ and $C_2$ in (5) are calculated to the 
next-to-leading order in reference\cite{buras} resulting in $a_2=C_2(\mu 
)+1/3C_1(\mu )=0.155$ for $\mu =m_b\approx 5$ GeV.  However, the 
branching ratio for the exclusive decay $B\to K\psi$ obtained from (4) 
requires $a_2$ to be roughly by a factor of two larger than the above 
value in order to agree with the experimental data\cite{pdg96}
$$
BR(B^+\to K^+\psi )=(0.101\pm 0.014)\% \; .
$$ 
This discrepancy between theoretical prediction and measurement could be 
due to two factors.  On one hand, the $\mu$-dependence of the Wilson 
coefficients which arises from short distance QCD results in a 
significant uncertainty in the calculated decay rate in the context of 
factorization.  In fact, phenomenologically, $a_2$ is treated as a free 
parameter to be determined from experiment\cite{a2}.  On the other hand, 
one may question the validity of the factorization assumption which 
allows to infer eq. (4) from the effective Hamiltonian (2).  In other 
words, the second term in (2) which is nonfactorizable could have a 
significant contribution to the matrix element\cite{kr}..  To disentangle 
these two factors and examine the factorization assumption, the ratio of 
the inclusive decays $R$ can serve as a crucial testing ground.  Aside 
from the cancellation of the Wilson coefficients in $R$, this ratio is 
also free from the nonperturbative hadronic uncertainties which is 
usually associated with the theoretical calculations of exclusive 
decays\cite{am}.

Using (4) and (7), one can calculate the decay rates $\Gamma (b\to s\psi 
(\eta_c))$:
\begin{equation}
\Gamma (b\to s\psi )=\displaystyle \frac{C^2f_\psi^2}{8\pi 
m_bm_\psi^2}
\begin{array}[t]{l} g(m_b,m_s,m_\psi )\\  
                 \displaystyle  \times \left [ 
m_b^2(m_b^2+m_\psi^2)-m_s^2(2m_b^2-m_\psi^2)+m_s^4-2m_\psi^4\right ] \; , 
\end{array} 
\end{equation}
\begin{equation}
\hskip -1.3cm \Gamma (b\to s\eta_c )=\displaystyle 
\frac{C^2f_{\eta_c}^2}{8\pi m_b} g(m_b,m_s,m_{\eta_c} )   \left [ 
{(m_b^2-m_s^2)}^2-m_{\eta_c}^2(m_b^2+m_s^2) \right ] \; , 
\end{equation}
where
\begin{equation}
g(x,y,z)={\left [ 
{(1-\frac{y^2}{x^2}-\frac{z^2}{x^2})}^2-4\frac{y^2z^2}{x^4}\right 
]}^{1/2}\; ,
\end{equation}
and $m_b$ and $m_s$ are $b$ and $s$ quark masses, respectively.  The 
inclusive decay rates $\Gamma (B\to X_s\psi (\eta_c ))$ are usually 
approximated by eqs. (8) and (9).  However, in this work we estimate the 
bound state corrections to this approximation by taking into accout the 
motion of the heavy $b$ quark inside the B meson.  We follow the ACCMM 
method \cite{accmm} which incorporates the 
bound state effect in semileptonic B decays by assuming a virtual $b$ 
quark inside B meson accomponied by an on-shell light quark.  In the 
meson rest frame, the energy-momentum conservation leads to the following 
relation for $b$ quark mass $W$:
\begin{equation}
W^2({\bf p})=m_B^2+m_q^2-2m_B\sqrt{{\bf p}^2+m_q^2} \; ,
\end{equation}
where $m_q$ is the light quark mass and ${\bf p}$ is the 3-momentum of 
the $b$ quark.  Following reference \cite{accmm}, we also consider a 
Gaussian momentum distribution for the Fermi motion of $b$ quark:
\begin{equation}
\displaystyle \phi ({\bf p})=\frac{4}{\sqrt{\pi}p_F^3}e^{-{\bf p}^2/p_F^2} 
\; . 
\end{equation}
The model parameter $p_F$ determines the distribution width, and is 
related to the average momentum $<p>$.

At this point we would like to remark on the consistency of the above 
model with heavy quark expansion.  Let us consider the average $b$ quark 
mass ${\overline m}_b$ defined as:
\begin{equation}
{\overline m}_b=\int_0^{ p_{\rm max}}W({\bf p})\phi ({\bf p})p^2dp\; ,
\end{equation}
where $p_{\rm max}$ is the maximum kinematically allowed momentum.  Using 
eq. (13), one can derive an expansion of the B meson mass $m_B$ in powers 
of ${\overline m}_b$ as follows:
\begin{equation}
m_B={\overline m}_b+\frac{2p_F}{\sqrt\pi}+\frac{3p_F^2}{4{\overline m}_b}+
O \left ( \frac{1}{{\overline m}_b^2} \right )\; ,
\end{equation}
in which $m_q=0$ is assumed.  A comparison of eq. (14) with the usual heavy 
quark expansion formula for heavy-light mesons, i.e.
\begin{equation}
m_M=m_Q+\bar\Lambda -\frac{\lambda_1 +d_M\lambda_2}{2m_Q}+O\left 
(\frac{1}{m_Q^2}\right )\; ,
\end{equation}
reveals that once ${\overline m}_b$ is identified with the mass of the 
heavy quark $m_Q$ in eq. (15), the nonperturbative parameters 
$\bar\Lambda$ and $\lambda_1$ of the heavy 
quark expansion and the model parameter $p_F$ are connected as follows:
\begin{equation}
\bar\Lambda=\frac{2p_F}{\sqrt\pi}\;\;\; ,\;\;\; \lambda_1=-\frac{3p_F^2}{2} 
\; . 
\end{equation}
The ACCMM model does not provide a corresponding term for the 
nonperturbative parameter $\lambda_2$ ($d_M=3,-1$ for psuedoscalar and 
vector mesons, respectively) which is due to the spin interactions.  This 
could be considered as a shortcoming of the model.  However, the 
constraint imposed by eq. (16), i.e. $\lambda_1=-3\pi /8{\bar\Lambda}^2$, 
is in reasonable agreement with quoted values for these parameters 
\cite{gklw}.  It is in this sense that we consider the above model to be 
consistent with heavy quark symmetries.

To incorporate the effects of the motion of the $b$ quark in the 
inclusive decays $B\to X_s\psi (\eta_c )$, we replace the $b$ quark mass 
$m_b$ in eqs. (8) and (9) with $W({\bf p})$ defined in eq. (11) and 
integrate over the kinematically allowed range of the $b$ quark momentum 
${\bf p}$, i.e.
\begin{equation}
\Gamma (B\to X_s\psi (\eta_c ))=\int_0^{p_{max}}\Gamma (b\to s\psi (\eta_c) 
)_{m_b= W({\bf p})}\phi({\bf p})p^2dp \; .
\end{equation}
As a result, the sensitivity to the heavy quark mass $m_b$ is replaced 
with the model parameter $p_F$ and the light quark mass $m_q$ 
dependence.  There are various determinations of $p_F$ from fits to 
semileptonic $B$ decays and also from comparison with the heavy quark 
effective theory approach.  In reference \cite{gklw}, $\bar\Lambda 
=0.55\pm 0.05$ GeV and $\lambda_1 =-0.35\pm 0.05$ GeV$^2$ have been 
extracted from CLEO data on inclusive semileptonic $B\to 
X\ell\bar\nu_\ell$ decay.  A comparison with eq. (16) leads to 
$p_F\approx 0.5$ GeV.  However, in reference \cite{hkn}, a smaller central 
value 
$p_F=0.27^{+0.22}_{-0.27}$ is obtained by an ACCMM model analysis of the 
ARGUS results for the lepton energy spectrum of $B\to 
X_c\ell\bar\nu_\ell$ decay.  In order to 
investigate the sensitivity of our estimates to the model parameter 
$p_F$, we present results for $p_F=0.3$ and $p_F=0.5$ GeV.  Using eq. 
(14), these values of $p_F$ correspond to ${\overline m}_b=4.92$ GeV 
and ${\overline m}_b=4.65$ GeV, respectively.    

Following the prescription of eq. (17) and introducing the notation
\begin{equation}
\Gamma \left (B\to X_s\psi (\eta_c)\right 
)=\frac{C^2f^2_{\psi(\eta_c)}}{8\pi}N_{\psi (\eta_c)}\; ,
\end{equation}
we obtain $N_\psi =7.64(8.14)$ GeV and $N_{\eta_c}=44.29(48.18)$ GeV$^3$ 
for $p_F=0.3$ GeV (the first number is calculated by taking the strange 
quark mass $m_s=0.55$ GeV and the second number in parenthesis is resulted 
from using $m_s=0.15$ GeV).  Comparing these results with the case where 
$m_b={\overline m}_b=4.92$ GeV is inserted in eqs. (8) and (9), i.e. 
$N_\psi =7.55(8.04)$ GeV and $N_{\eta_c}=43.55(47.42)$ GeV$^3$, indicates 
that only a small correction of order $\% 1-2$ arises from considering 
this bound state effect.  On the other hand, a larger value of the model 
parameter $p_F=0.5$ GeV (which is compatible with heavy quark effective 
theory approach) results in $N_\psi =5.85(6.35)$ GeV and 
$N_{\eta_c}=33.14(36.77)$ GeV$^3$.  A comparison with the decay rates 
obtained from eqs. (8) and (9) by using $m_b={\overline m}_b=4.65$ GeV, i.e. 
$N_\psi =5.57(6.06)$ GeV and $N_{\eta_c}=31.01(34.62)$ GeV$^3$, reveals a 
larger bound state corrections of order $\% 5-6$.  We also notice that 
the corrections are almost independent of $s$ quark mass $m_s$.   

As we observe from eq. (18), the ratio $R=\Gamma (B\to X_s\eta_c)/\Gamma 
(B\to X_s\psi )$ is independent of $C$ in the context of factorization 
assumption, i.e. 
\begin{equation}
R=\frac{\Gamma (B\to X_s\eta_c)}{\Gamma (B\to X_s\psi )}=
\frac{N_{\eta_c}f^2_{\eta_c}}{N_\psi f^2_\psi}
=(5.8\pm 0.1\; {\rm GeV}^2) 
\frac{f^2_{\eta_c}}{f^2_\psi}\; ,
\end{equation}
where our error estimate in the numerical factor represents the variation of 
the 
model parameter $p_F$ in the range $0.3$ to $0.5$ GeV and also the $s$ 
quark mass $m_s$ from $0.15$ to $0.55$ GeV.

To evaluate the ratio of the decay constants in (19), we use the 
potential model relations which relate these form factors to the value of 
the meson wavefunction at the origin:
\begin{equation}
\begin{array}{l}
\displaystyle f_{\eta_c}=\sqrt{\frac{12}{m_{\eta_c}}}\Psi_{\eta_c}(0)\; , \cr
\displaystyle f_\psi=\sqrt{12m_\psi}\Psi_\psi (0)\; .
\end{array}
\end{equation}
The common assumption in the literature is that the wavefunction of the 
psuedoscalar and vector mesons at the origin are more or less identical.  
However, in reference \cite{prd95}, based on a simple perturbation theory 
argument, this ratio was estimated to be 
\begin{equation}
\frac{{\vert \Psi_{\eta_c}(0)\vert}^2}
{{\vert \Psi_\psi(0)\vert}^2}
=1.4\pm 0.1\; .
\end{equation}
Consequently, the ratio of the decay constants in (19) can be written as:
\begin{equation}
\frac{f^2_{\eta_c}}{f^2_\psi}=
\frac{1}{m_{\eta_c}m_\psi}
\frac{{\vert \Psi_{\eta_c}(0)\vert}^2}
{{\vert \Psi_\psi(0)\vert}^2}
\approx 0.15\pm 0.01 \;\; {\rm GeV}^{-2}\; .
\end{equation}
Inserting (22) into (19), we obtain:
\begin{equation}
R=\frac{\Gamma (B\to X_s\eta_c)}{\Gamma (B\to X_s\psi )}=0.87\pm 0.06\; .
\end{equation}
We would like to point out again that the ratio $R$ is free of hadronic 
model 
uncertainties which are normally encountered in calculating exclusive 
decays.  The inclusive decay branching ratio $BR(B\to X_s\psi )=(1\pm 
0.25)\times 10^{-2}$ has been extracted from experimental data for direct 
production of $\psi$ in nonleptonic $B$ decays \cite{dtp}.  Therefore, a 
measurement of the inclusive $B\to X_s\eta_c$ decay along with reducing 
the error bar in the experimental result for $B\to X_s\psi$ can serve as 
an alternative test of the factorization assumption which is used in 
deriving eq. (23).  A significant deviation of the experimental value for 
$R$ from the theoretical prediction in eq. (23) could be an indication of 
the failure of the factorization assumption.  We would like to emphasize 
that the cancellation of the Wilson coefficients in the ratio $R$ results 
in a significant reduction of the uncertainty due to QCD corrections and 
scale dependence.

In conclusion, using factorization, we calculated the ratio of the 
inclusive nonleptonic $B$ decays to a hadron containing a strange quark 
plus psuedoscalar and vector $c\bar c$ mesons.  The bound state effect 
due to the motion of the $b$ quark inside the $B$ meson was also 
considered.  
We used an improved estimate of the ratio of the dacay constants for 
$\eta_c$ and $\psi$ was obtained 
based on potential model and pertubation theory argument.  Once the 
experimental results on $B\to X_s\eta_c$ decay are available, a 
comparison to the theoretical prediction presented in this paper could 
serve as a test of factorization assumption.  Finally, we would like to 
emphasize that the pattern of deviations from factorization can give us 
important clues on the nonfactorizable contributions, and therefore, it 
is important to examine the departures from this approximation using 
various experimental data.

\vskip 1.0cm
{\bf \large Acknowledgement}

The authors would like to thank A. Ali and T. Morozumi for useful 
discussions.  M. A.'s work is supported by the Japanese Society for the 
Promotion of Science.

\end{document}